\documentclass[fleqn,usenatbib]{mnras}

\usepackage{newtxtext,newtxmath}
\usepackage[T1]{fontenc}
\usepackage{ae,aecompl}
\usepackage{ulem} 

\usepackage{graphicx}	% Including figure files
\usepackage{amsmath}	% Advanced maths commands
\usepackage{amssymb}	% Extra maths symbols
\hypersetup{draft}
\title[Surface density: a new parameter in the fundamental metallicity relation of star-forming galaxies]{Surface density: a new parameter in the fundamental metallicity relation of star-forming galaxies}

\author[Hashimoto et al.]{
Tetsuya Hashimoto,$^{1}$\thanks{E-mail: tetsuya@phys.nthu.edu.tw}
Tomotsugu Goto,$^{1}$ and Rieko Momose $^{1}$
\\
$^{1}$Institute of Astronomy, National Tsing Hua University, 101, Section 2. Kuang-Fu Road, Hsinchu, 30013, Taiwan (R.O.C.)\\
}

\date{Accepted 2018 January 10. Received 2018 January 10; in original form 2017 August 6}

\pubyear{2018}

\begin{document}
\label{firstpage}
\pagerange{\pageref{firstpage}--\pageref{lastpage}}

\maketitle

\begin{abstract}
Star-forming galaxies display a close relation among stellar mass, metallicity and star-formation rate (or molecular-gas mass).
This is known as the fundamental metallicity relation (FMR) (or molecular-gas FMR), and it has a profound implication on models of galaxy evolution. 
However, there still remains a significant residual scatter around the FMR. 
We show here that a fourth parameter, the surface density of stellar mass, reduces the dispersion around the molecular-gas FMR. 
In a principal component analysis of 29 physical parameters of 41,338 star-forming galaxies, the surface density of stellar mass is found to be the fourth most important parameter. 
The new four-dimensional fundamental relation forms a tighter hypersurface that reduces the metallicity dispersion to 50\% of that of the molecular-gas FMR. 
We suggest that future analyses and models of galaxy evolution should consider the FMR in a four-dimensional space that includes surface density. 
The dilution time scale of gas inflow and the star-formation efficiency could explain the observational dependence on surface density of stellar mass.
\end{abstract}


%%%%%%%%%%%%%%%%%%%%%%%%%%%%%%%%%%%%%%%%%%%%%%%%%%

%%%%%%%%%%%%%%%%% BODY OF PAPER %%%%%%%%%%%%%%%%%%

\section{Introduction}

The relation between stellar mass and gas-phase metallicity, which was discovered a decade ago, is one of the most fundamental relations for star-forming galaxies both in the local universe and at high redshifts \citep{Tremonti2004,Erb2006}.
The gas-phase metallicity, here, is a measure of the relative amount of oxygen and hydrogen present.
Many explanations for this relation have been proposed, including metal depletion by enriched gas outflow, which depends on the galactic gravitational well \citep{Garnett2002,Tremonti2004,Koppen2007,Brooks2007,Dalcanton2007,Calura2009,Finlator2008}. 
Since its discovery, this relation has been developed into what is known as the fundamental metallicity relation (FMR) among stellar mass, metallicity and star-formation rate (SFR), defining a curved surface in the equivalent three-dimensional (3D) parameter space \citep{Mannucci2010}. 
In \citet{Mannucci2010}, the large dispersion around the stellar-mass metallicity relation can be explained by the difference of SFRs of individual galaxies, which reduces the scatter significantly.
Recent instances of the FMR based on large statistical samples of star-forming galaxies have dramatically advanced our understanding of galaxy evolution.
For example, the negative correlation that exists between SFR and metallicity in the FMR can be understood in the context of the inflow of metal-poor gas that directly dilutes the gas-phase metallicity and also triggers star formation \citep{Mannucci2010,Forbes2014}. 
The observed strength of the negative correlation between SFR and metallicity may depend on different metallicity calibration methods \citep{Telford2016,Kashino2016} and different ways to be observed, i.e. single-fiber or integral-field observations \citep{Sanchez2017,Barrera-Ballesteros2017}.

There remains a significant residual scatter in metallicity beyond measurement error even after the SFR dependence is removed \citep{Salim2014,Salim2015}.
This suggests that there could be an other factor which regulates metallicity of star-forming galaxies except for SFR.
The possible fourth parameter following SFR could reduce the scatter around the 3D relation as the third parameter, SFR, reduced the dispersion around the mass-metallicity relation.
The goal of this paper is finding out the fourth most important physical parameter of star-forming galaxies which reduces the scatter around the 3D relation.
We describe the sample selection and present a brief summary of 29 physical parameters collected for individual galaxies in section \ref{sample_selection}.
In section \ref{PCA}, we explain how to select candidates among 29 parameters.
We examine actual tightnesses of fundamental relations defined by various combinations of candidate parameters in section \ref{median surface}.
Our finding of the possible fourth most important parameter is shown in section \ref{surface_density}.
We present a comparison of the resulting metallicity dispersions around the fundamental relations with previous works in section \ref{disp_metal}.
In section \ref{discussion} we discuss possible physical interpretations of the fourth parameter.
We finally summarize and conclude in section \ref{conclusion}.

\section{Sample selection}
\label{sample_selection}
The galaxies in our sample were selected from the Sloan Digital Sky Survey (SDSS) Data Release 7 \citep{Abazajian2009}.
We used the publicly available catalogue of emission-line fluxes produced in collaboration between the Max Planck Institute for Astrophysics (MPA) and Johns Hopkins University (JHU) \citep{Kauffmann2003,Brinchmann2004,Salim2007}.
The criteria for selecting star-forming galaxies were based on previous research on the stellar-mass-metallicity relation \citep{Kewley2008}.
A signal-to-noise ratio (S/N) of at least 8 was required in the strong optical emission lines of [O~{\sc ii}]$\lambda$3726, 3729, H$\beta$, [O~{\sc iii}]$\lambda$5007, H$\alpha$, [N~{\sc ii}]$\lambda$6584 and [S~{\sc ii}]$\lambda$6717, 6731 for reliable estimates of metallicity and electron density.
The median values of S/N of these emission lines in our sample are 15.5, 17.5, 32.7, 18.0, 81.5, 41.3, 28.4, and 21.4, respectively.
This S/N limit on emission lines ensures accurate measurements of physical parameters but could lead to a biased sample.
Galaxies with strong emission lines could be biased towards higher SFR and lower metallicity, which would enhance the dependence of metallicity on SFR \citep{Reyes2015}.
This is probably because the emission-line fluxes are directly related to SFR and metallicity.
Our present target of surface density of stellar mass was estimated from continuum fluxes, and it is independent of the emission lines.
Therefore, the S/N criteria are unlikely to have an impact on our analysis, although the parameter range of metallicity may be biased to some extent.
The redshift range was limited to 0.04-0.1.
The lower redshift limit ensured at least 20\% higher fibre covering fraction of the total photometric $g'$-band light as required for metallicity to approximate the global value \citep{Kewley2005}.
The upper redshift limit is due to the fact that the SDSS sample of star-forming galaxies becomes incomplete at redshifts above 0.1 \citep{Kewley2006}.
Host galaxies of active galactic nuclei (AGNs) were excluded by using the emission-line-ratio diagnostic diagram (i.e. a Baldwin-Phillips-Terlevich diagram \citep{Baldwin1981}) and the AGN selection criteria \citep{Kauffmann2003}.
The resulting sample contained 41,338 star-forming galaxies after removing duplicates in the emission-line catalogue.
In summary, we used the following criteria to select star-forming galaxies.\\

\begin{itemize}
\item S/N $>$ 8 for strong optical emission lines
\item 0.04 $<$ redshift $<$ 0.1
\item $\log$([O~{\sc iii}]$\lambda$5007/H$\beta$) $<$ 0.61/($\log$([N~{\sc ii}]$\lambda$6584/H$\alpha$)-0.05) + 1.3
\end{itemize}

In total, 18 physical parameters of the sampled galaxies were compiled from various literatures, and a further 11 were calculated in this work.
Note that 11 parameters include N2O3 index which is defined as N2O3 $=\log$[([N~{\sc ii}]$\lambda 6584/{\rm H}\alpha$)/([O~{\sc iii}]$\lambda 5007/{\rm H}\beta$)].
This emission-line ratio is linked to metallicity according to the calibration formula of 12+log(O/H)=8.73+0.32 $\times$ N2O3 with a statistical uncertainty of $\sim$ 0.1 dex \citep{Pettini2004, Marino2013}.
Throughout this paper, we use N2O3 as an indicator of metallicity to avoid this large statistical uncertainty which is involved when N2O3 is converted to the actual value of metallicity.

We also note that 29 parameters contain not only single measurements but also their combinations such as molecular-gas mass which can be calculated from SFR and half-light radius assuming the Kennicutt-Schmidt law \citep{Kennicutt1998}.
The molecular-gas mass is mathematically equivalent to the combination of the SFR and half-light radius here.
However the physical meaning of molecular-gas mass is different from SFR and half-light radius actually.
It is possible for molecular-gas mass to be more important than SFR (or radius) itself according to underlying physical mechanism to regulate other parameters such as metallicity.
Therefore we treat combinations as independent parameters.

The 29 parameters are summarised in Table \ref{tab1} and details are in APPENDIX A.

\section{Principal component analysis}
\label{PCA}
Principal component analysis (PCA) is a useful method for investigating the relative contributions of individual parameters to the observed dispersion in a set of data. 
It involves constructing a new set of parametric axes PC1, PC2, ..., PC$n$ that are ordered by decreasing contribution to the total dispersion and that are formed from linear combinations of the original parameters. 
These new axes are mutually perpendicular and not to change the original data distribution. 
To find a new fundamental parameter for the FMR, we carried out PCA on 41,338 star-forming galaxies selected from the SDSS Data Release 7 \citep{Abazajian2009}. 
We included 29 physical parameters categorised into six groups, namely, MASS, METAL, ACTIVITY, SIZE/MORPHOLOGY, ENVIRONMENT and OTHER (see Table \ref{tab1} and APPENDIX A).
The resulting \lq factor loadings\rq are indicators of how strongly each original parameter contributes to the new axes (see APPENDIX B).
The large absolute value of the factor loading indicates large contribution to the new PC axes.
The same signs of factor loadings mean positive correlation between the parameters.
Different signs show anti-correlation between the parameters.

\begin{figure*}
	\includegraphics[width=15cm]{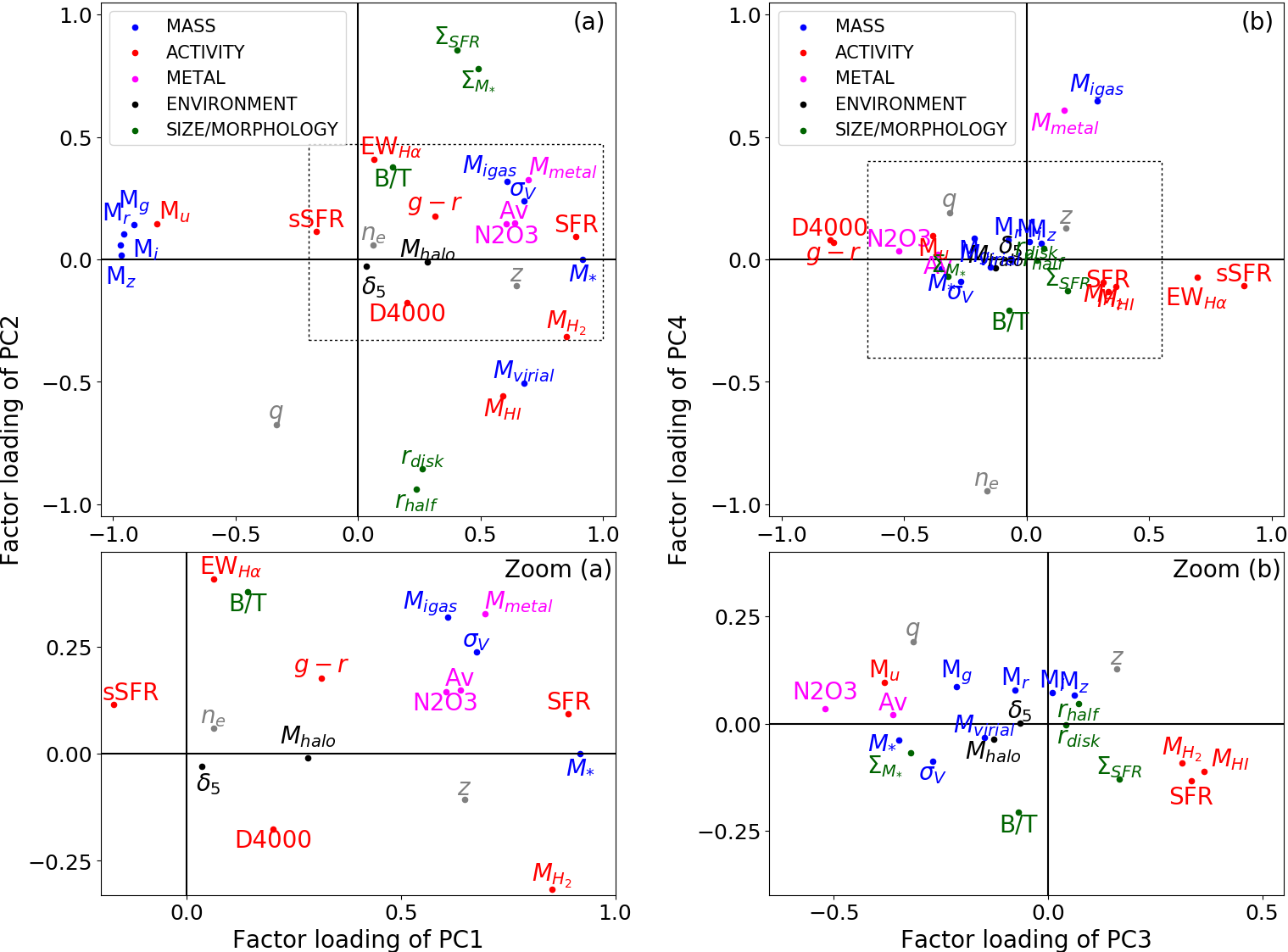}
    \caption{
    Factor loadings of principal components PC1, PC2, PC3 and PC4.
    The square of factor loading is an indicator of how strongly any original parameter contributes to the principal components. 
    The same signs of factor loading means positive correlation between the parameters.
    Different signs show anti-correlation between the parameters.
    Colours show groupings of 29 parameters of star-forming galaxies. 
    Dashed rectangular regions in the upper panels are magnified in the lower panels. 
    The parameters are summarised in Table \ref{tab1} (see APPENDIX A for details).
    }
    \label{fig1}
\end{figure*}

In particular, the factor loadings of the parameters in the MASS, ACTIVITY and METAL groups show relatively high absolute values along the PC1 axis (Figure \ref{fig1}a).
The archetypes of these three groups (i.e. stellar mass, SFR and N2O3, respectively) have the same sign along PC1, reflecting the well-known first-order stellar-mass-SFR and stellar-mass-metallicity relations. 
The main contributors to PC2 are the SIZE/MORPHOLOGY parameters.
As for PC3, the ACTIVITY and METAL parameters show higher absolute values of factor loadings compared to those in the MASS group (Figure \ref{fig1}b).
The factor loading of N2O3 is opposite in sign to the indicators of star formation, which suggests a general trend of negative correlation between the star-forming activity and metallicity such as the SFR-metallicity anti-correlation reported by \citet{Mannucci2010}.

The aforementioned correlations, to some extent, represent known relations in the FMR \citep{Mannucci2010}.
However, the distribution of 29 parameters in Figure \ref{fig1} shows complex inter-relations between many parameters. 
Thus quantitative interpretations of individual PC axes are quite complicated.

To avoid the complexity in Figure \ref{fig1}, we adopt a criteria to select important parameters from this diversity based on the distance in the 3D space of factor loadings of PC1, PC2 and PC3 (instead of using values along each PCA axis).
The PC1, PC2 and PC3 axes explain most ($\sim$70\%) of the overall distribution of galaxies, suggesting the contribution from the fourth parameter to these three axes if it exists.
Therefore the comprehensive contributions of the original parameters to the galactic distribution are indicated by the radial distances from the origin in the 3D space of the PC1, PC2 and PC3 factor loadings.
The physical parameters in each categories are listed in Table \ref{tab1} in order of the distance in PC1-3 factor loading space.
The most important parameters in each group (i.e. the most distant parameters in the 3D factor-loading space) are stellar mass (MASS), N2O3 (METAL), molecular-gas mass (ACTIVITY), surface density of stellar mass (SIZE/MORPHOLOGY), dark-matter halo mass (ENVIRONMENT) and ionisation parameter (OTHERS).
We conduct further exploration of these parameters as well as SFR and half-light radius, the latter two of which are included as existing references. 

Of the remaining parameters, we note electron density, ionised-gas mass and metal mass as being outliers along PC4, which were calculated by using electron density (see APPENDIX A).
However, unlike the parameters along PC3, the aforementioned three parameters are widely spread, implying a little correlation with other parameters such as stellar mass, N2O3, molecular-gas mass and surface density of stellar mass.

\begin{table*}
	\centering
	\caption{
    Summary of 29 parameters of SDSS star-forming galaxies, categorised into six groups (see APPENDIX A for details) in order of radial distance in the 3D space of the factor loadings of PC1, PC2 and PC3.
    }
	\label{tab1}
      \begin{tabular}{|ll|ll|ll|}\hline
\multicolumn{2}{|c|}{MASS} & \multicolumn{2}{|c|}{METAL} & \multicolumn{2}{|c|}{ACTIVITY} \\ \hline
$M_{*}$& Stellar mass & N2O3& Metallicity indicator & $M_{\rm H_{2}}$& Molecular-gass mass  \\
M$_{i}$& $i$-band absolute magnitude & $M_{metal}$& Metal mass in ionised gas & SFR& Star-formation rate \\
M$_{z}$& $z$-band absolute magnitude & Av& Dust extinction & M$_{u}$& $u$-band absolute magnitude \\
M$_{r}$& $r$-band absolute magnitude & & & sSFR& Specific star-formation rate \\
M$_{g}$& $g$-band absolute magnitude & & & $M_{\rm HI}$& Atomic hydrogen gas mass  \\
$M_{virial}$& Virial mass ($2r_{half}\sigma^{2}/G$) & & & g-r& Colour \\
$\sigma$& Velocity dispersion & & & D4000& Strength of 4000-\AA\ break \\
$M_{igas}$& Ionised-gas mass & & & EW$_{\rm H\alpha}$& Equivalent width of H$\alpha$ \\\hline \hline
\multicolumn{2}{|c|}{SIZE/MORPHOLOGY} & \multicolumn{2}{|c|}{ENVIRONMENT} & \multicolumn{2}{|c|}{OTHER} \\ \hline
$\Sigma_{M_{*}}$& Surface density of M$_{*}$ & $M_{halo}$& Dark-matter halo mass & q& [O~{\sc iii}]$\lambda$5007/[O~{\sc ii}]$\lambda$3727 \\
$r_{half}$& $r$-band half-light radius & $\delta_{5}$& Local galaxy number density & $z$& Redshift \\
$\Sigma_{\rm SFR}$& Surface density of SFR & & & $n_{e}$& Electron density \\
$r_{disk}$& $r$-band disk radius & & & & \\
B/T& Bulge-to-total fraction & & & & \\ \hline
  \end{tabular}
\end{table*}

\section{3D/4D median surfaces of fundamental relations}
\label{median surface}
In section \ref{PCA}, stellar mass, N2O3, molecular-gas mass, surface density of stellar mass, dark-matter halo mass and ionization parameter are selected as candidates of fundamental physical parameters of star-forming galaxies.
In this section we calculate tightnesses of fundamental relations defined by various combinations of these candidate parameters to find out the fourth most important parameter which reduces the scatter around the 3D fundamental relation.
We here performed additional PCAs for various 3D combinations of candidate parameters (not 29 in section \ref{PCA}) to carefully define the representative surface distributions of data points.
Note that PCAs here are independent of the analysis in section \ref{PCA} and performed only for the purpose of calculation of the representative surface in 3D space.

The data points are mostly spread out in the PC1-PC2 space because PC3 makes the smallest contribution to the total dispersion.
This means that the median values along PC3 reflect the representative curved surface distribution of the data points.
Therefore, to derive the 3D fundamental relation, we calculated the median distribution of data points along the PC3 in the PC parameter space.
The PC1-PC2 space is split into grids in units of $\sigma_{\rm PC1}$ $\times$ $\sigma_{\rm PC2}$, where $\sigma_{\rm PC1}$ and $\sigma_{\rm PC2}$ are the dispersions of the data points along each axis.
A unit of five grid neighbours is used to calculate a local median surface. 
The offset of each data point is defined by the orthogonal distance from the closest median surface.
We extended the same analysis to the four-dimensional (4D) fundamental relation by adding an extra dimension to the 3D one.

\section{Surface density-N2O3 relation}
\label{surface_density}
\begin{figure*}
	\includegraphics[width=15cm]{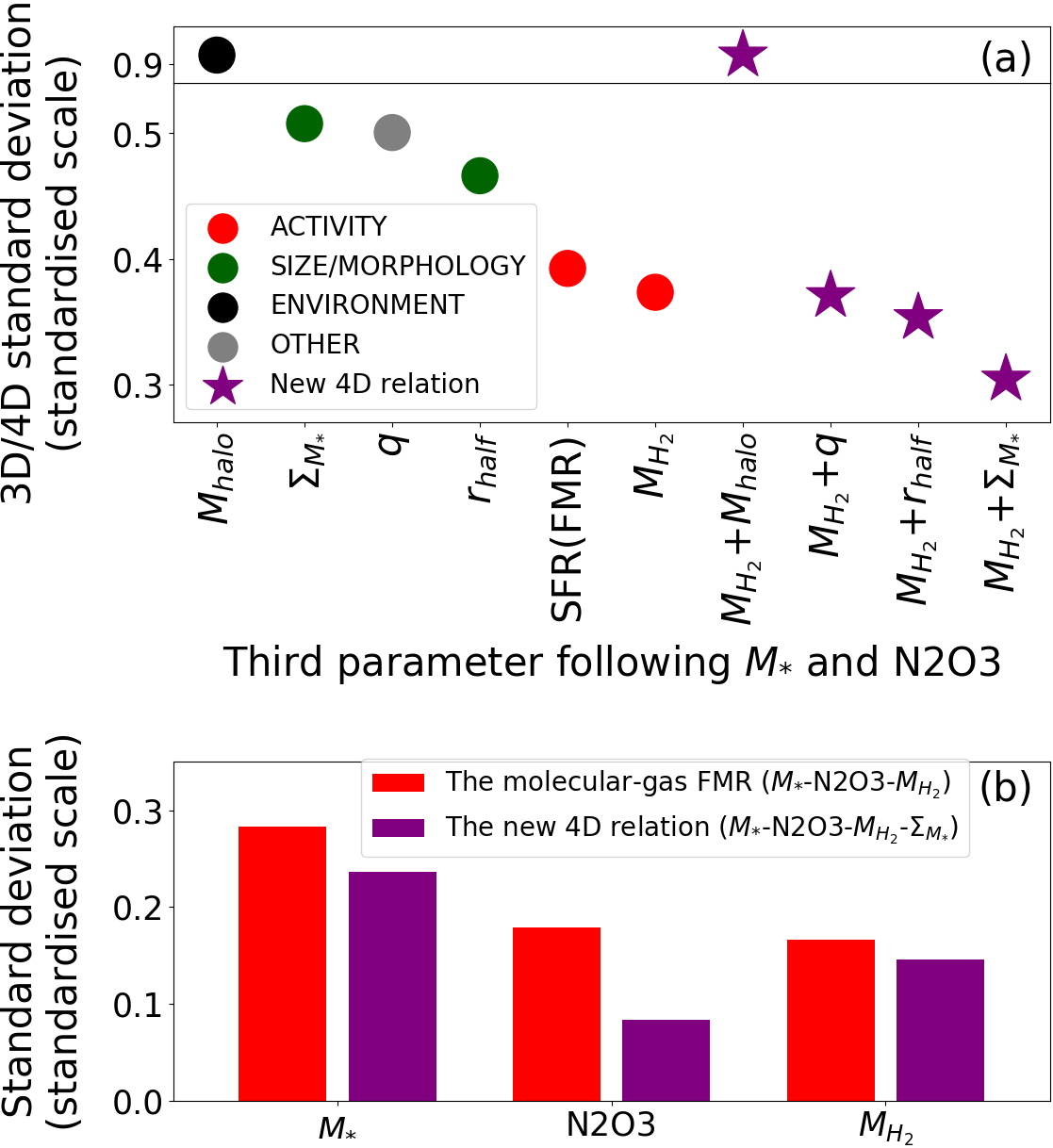}
    \caption{
    Standard deviations around fundamental relations in the standardised scale.
    \textbf{(a),} Standard deviations around fundamental relations in 3D parameter spaces defined by stellar mass, N2O3 and various third parameters (red, green, black and grey dots), also including 4D parameter spaces (purple stars).
    \textbf{(b),} Breakdown of standard deviations around the 3D and 4D relations (i.e. the molecular-gas FMR and the addition of the surface density of stellar mass) into triaxial components of stellar mass, N2O3 and molecular-gas mass.
    }
    \label{fig2}
\end{figure*}

Figure \ref{fig2}a shows the standard deviations around the median surfaces of the fundamental galactic relations defined by stellar mass, N2O3 and various choices for the third parameter.
The standard deviations were calculated using orthogonal distances from the median surfaces in the standardised 3D parameter spaces so that the tightness of the different relations could be fairly compared with each other.
A set of more fundamental parameters should construct a tighter relation according to the underlying physical mechanism of star-forming galaxies.
The use of SFR as the third parameter corresponds to the conventional FMR and shows the second smallest scatter among the various choices for the third parameter (coloured dots in Figure \ref{fig2}a).

Recently, it has been suggested that molecular-gas mass is more fundamental than SFR as the third parameter (the former option is referred to hereinafter as the molecular-gas FMR) because of the large contribution of the molecular-gas mass to metallicity that is indicated from the fundamental plane of the galactic distribution \citep{Bothwell2013,Bothwell2016a,Bothwell2016b}.
Although it is beyond the scope of this paper to judge which is more important as the third parameter (SFR or molecular-gas mass), we attempted to estimate the molecular-gas mass by assuming the well-established Kennicutt-Schmidt law \citep{Kennicutt1998} that connects SFR and molecular-gas mass (see APPENDIX A).
This is an indirect way to estimate the molecular-gas mass, but the alternative option of estimating it directly by CO observation is time-consuming and anyway is not feasible for the large sample used here (furthermore, the CO conversion factors are uncertain and could depend on metallicity \citep{Wolfire2010,Feldmann2012,Schruba2012}, which we wish to predict).
In Figure \ref{fig2}a, the molecular-gas FMR actually results in slightly less scatter than does the conventional FMR, which may support the use of molecular-gas mass as the third fundamental parameter. 

The standard deviation of the 3D relation defined by stellar mass, N2O3 and surface density of stellar mass (left green dot in Figure \ref{fig2}a) is not as small as that by using molecular-gas mass, confirming that molecular-gas mass is the third parameter. 
However, if surface density of stellar mass is incorporated into the molecular-gas FMR as an additional fourth parameter, the extent of galactic dispersion is indeed reduced (rightmost purple star in Figure \ref{fig2}a). 
The standard deviations around the molecular-gas FMR and this new 4D relation including surface density of stellar mass are separated into the triaxial components of stellar mass, N2O3 and molecular-gas mass in the standardised scale in Figure \ref{fig2}b.
The reduction in dispersion by adding the fourth parameter is predominant in N2O3; the reductions in the other parameters are more moderate. 

\begin{figure*}
	\includegraphics[width=15cm]{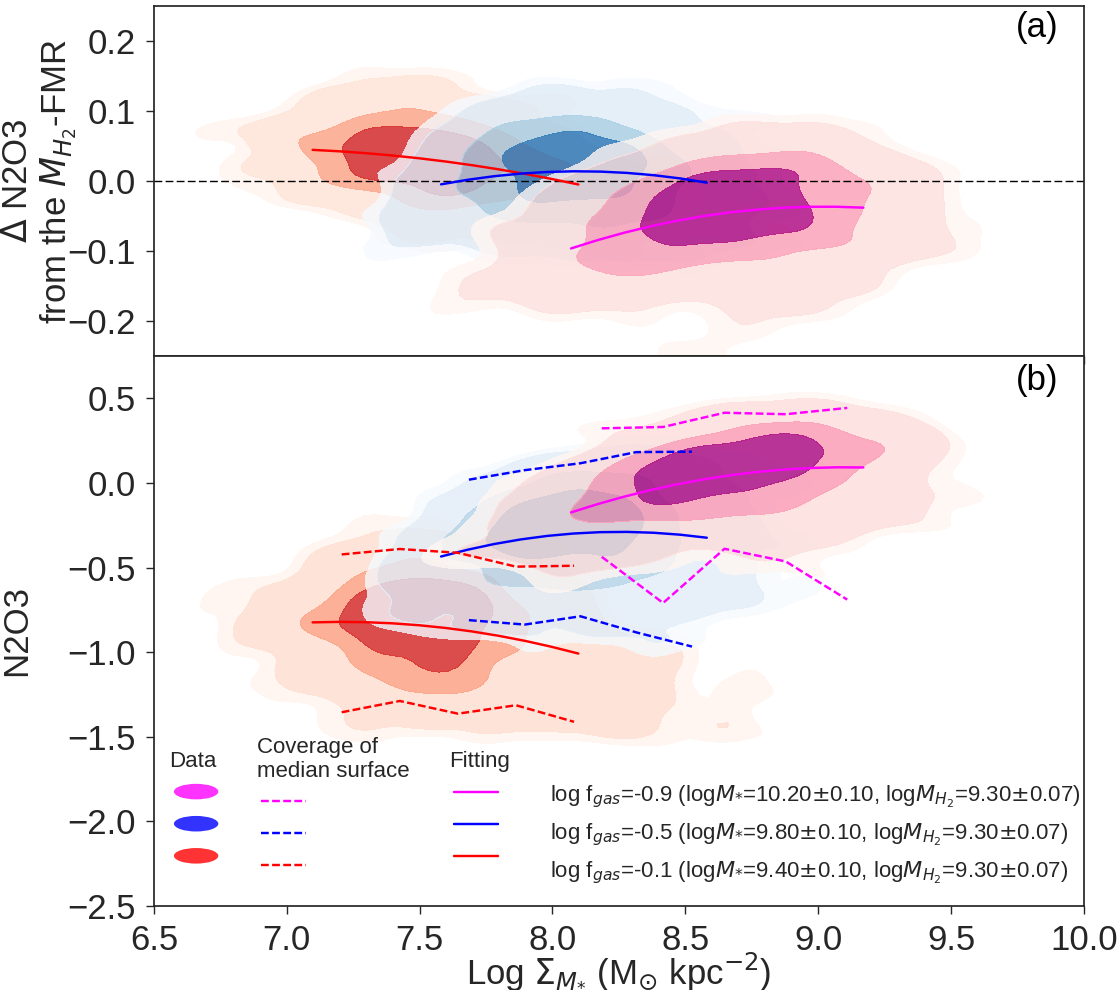}
    \caption{
    N2O3 dependency on the surface density of stellar mass in the physical scale.
    \textbf{(a),} N2O3 offset around the molecular-gas FMR, i.e. the 3D relation of stellar mass, N2O3 and molecular-gas mass as a function of surface density of stellar mass.
    The observed data are sampled from gas-mass-fraction bins.
    Galaxies with three different gas mass fractions are displayed in magenta, blue and red.
    Contours contain 38.2\%, 68.3\%, 95.5\% and 99.7\% of each binned sample.
    Solid lines are polynomial fits of each binned sample.
    Horizontal dashed line corresponds to the surface of the molecular-gas FMR defined by the median distribution of data points in the 3D space.
    \textbf{(b),} Same as (a) but for N2O3 in the vertical axis.
    The two dashed lines of each binned sample indicate the coverage of the projected 4D fundamental relation defined by median data points within the binning size of the sampled data.
    The lower and upper dashed lines are 1 and 99 percentiles of the N2O3 of the sampled galaxies expected from the 4D fundamental relation.
    }
    \label{fig3}
\end{figure*}

We suggest the correlation between surface density of stellar mass and N2O3 as the primary cause of the residual scatter around the molecular-gas FMR. 
This is confirmed clearly in Figure \ref{fig3}a, where we show the projected N2O3 offsets from the molecular-gas FMR as a function of surface density of stellar mass.
Here the molecular-gas FMR means the median surface of the data distribution in the 3D space of stellar-mass, N2O3 and molecular-gas mass, which is one of calculations described in section \ref{median surface}.
The vertical axis is the orthogonal offset from the molecular-gas FMR projected on N2O3 axis.

Note that hereinafter all physical parameters are expressed in physical scales for reference to previous studies unless otherwise noted.
In Figure \ref{fig3}a, galaxies are sampled into three different stellar-mass bins and a fixed molecular-gas mass (i.e. different gas mass fractions $f_{gas}=M_{H_{2}}/M_{*}$) to minimise the effects of stellar mass and molecular-gas mass on N2O3.
High-$f_{gas}$ galaxies show a negative correlation between the surface density of stellar mass and N2O3 offset, whereas low-$f_{gas}$ galaxies show a positive relation within the N2O3 dispersion around the molecular-gas FMR. 
The vertical dynamic ranges of these correlation/anti-correlation are larger than the expected uncertainty of N2O3 offset measurements in our sample ($\sim$0.04 dex discussed in section \ref{disp_metal}).

N2O3 (not offset) as a function of surface density of stellar mass is shown in Figure \ref{fig3}b for the three galaxy samples.
According to the molecular-gas FMR, the N2O3 of the samples should not show any surface-density dependence because of the fixed stellar mass and molecular-gas mass.

\begin{figure*}
	\includegraphics[width=15cm]{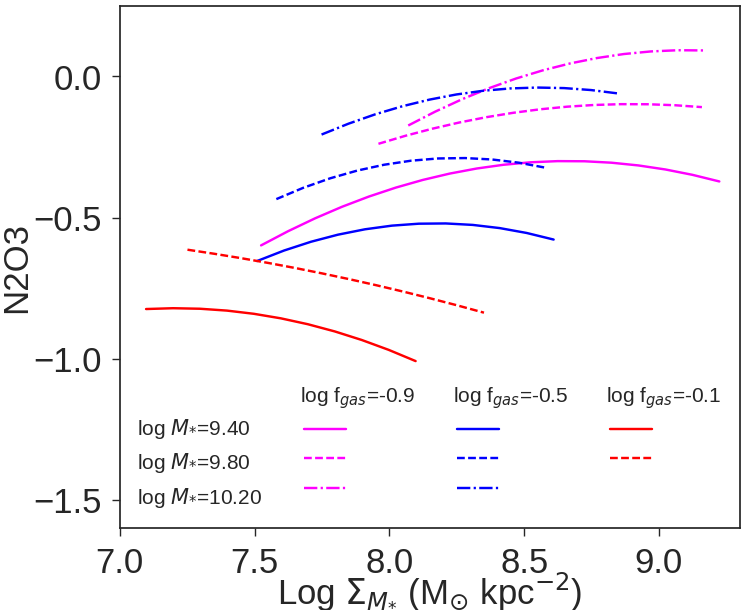}
    \caption{
    Dependency of the relation between surface density of stellar mass and N2O3 on gas-mass fraction.
    Galaxies are divided into eight-bin samples with different stellar masses and gas mass fractions.
    Polynomial fits of the eight-bin samples are shown as lines with different colours and styles, which correspond to the different gas mass fractions and stellar masses, respectively.
    }
    \label{fig4}
\end{figure*}

However, in Figure \ref{fig3}b, the three samples display the same trend, namely, a negative relation for high-$f_{gas}$ galaxies and a positive relation of low-$f_{gas}$ galaxies, which is consistent with Figure \ref{fig3}a.
In Figure \ref{fig3}, it is not still clear whether stellar mass governs the relation, or $f_{gas}$ does.
To clarify this point, in Figure \ref{fig4}, we changed the $f_{gas}$ with fixed stellar mass.
These trends are also confirmed in Figure \ref{fig4}, which shows a more detailed breakdown of the galaxies into eight bins. 
Although there is no polynomial fit for the massive/high-$f_{gas}$ bin (bottom right in the key) because the sample was too small, galaxies in the same range of stellar mass (the same line style) display different relations depending on $f_{gas}$.
For the same $f_{gas}$ (the same colour), galaxies with different stellar masses show similar dependencies. 
Therefore, the general trend of each surface-density-N2O3 relation depends on $f_{gas}$ rather than on stellar mass or molecular-gas mass.
These correlations underlying the molecular-gas FMR contribute to the dispersion around the 3D relation.

In relation to the fourth parameter, pioneering studies suggest a dependence on size of galaxy in the mass-metallicity relation (not the FMR) \citep{Ellison2008, Brisbin2012, Yabe2014, Salim2014}.
A 4D relation among stellar mass, metallicity, SFR and ionisation parameter has also been reported \citep{Nakajima2014}.
Figure \ref{fig1} confirms that these parameters show relatively high absolute values of factor loading. 
However, these values are still lower than that of surface density of stellar mass.
In fact, as shown in Figure \ref{fig2}, the size and ionisation parameters do not construct the tightest relation in either 3D or 4D, which suggests that surface density of stellar mass is more fundamental. 
The present work is the first time that the relative dependence of the full FMR is shown quantitatively.

\section{N2O3 dispersion around the 3D/4D relations}
\label{disp_metal}
\subsection{Observed N2O3 dispersion}
\label{disp_metal_obs}
It is well-known that different metallicity-calibration methods lead to different absolute values of metallicity \citep{Kewley2008}, which causes a large systematic error.
A relative value of metallicity (e.g. dispersion or offset) can be measured more accurately as long as the same method is used to calibrate the metallicity. 
The metallicity calibrations themselves have large statistical uncertainties, e.g. $\sim$0.1 dex for PP04 method \citep{Pettini2004,Marino2013}. 
Thus uncertainty of $\sim$0.1 dex is always involved when the emission-line ratio is converted to the metallicity.
As long as N2O3 is adopted instead of the conversion to metallicity unit, we lose the information on actual values of metallicity of individual galaxies.
However N2O3 still works statistically as an indicator of metallicity when we collect a large amount of sample.

The 3D standard deviation of the molecular-gas FMR corresponds to 0.07 dex in the projected N2O3 axis in the physical scale.
The dispersion of 0.07 dex in N2O3 corresponds to 0.023 dex in PP04 metallicity calibration \citep{Pettini2004}, which is a much smaller deviation than those in previous works, i.e., $\sim$0.05 dex \citep{Mannucci2010,Telford2016,Sanchez2017,Barrera-Ballesteros2017}.
This is mainly due to the different method used here to estimate the dispersion around the relation.
The orthogonal distance adopted here provides more reliable information on the tightness of each relation than does the distance along only the metallicity axis as used in previous works.
In general, the projection of orthogonal distance is smaller than the distance calculated along an arbitrary axis.
As a pure geometric effect, tilt of data distribution reduces the projected orthogonal distance by a factor of (1-cos$^{2}\theta$) as a first-order approximation.
The mass-N2O3 relation shows roughly $\theta$=45 degree in the standardized space which corresponds to factor of $\sim$0.5 (see also Figure \ref{fig7}).
Metallicity dispersion in our analysis divided by 0.5 is 0.023/0.5 = 0.046 dex.
This value is almost comparable to results of other works ($\sim$0.05 dex).

Figure \ref{fig5} shows the peaked distribution of N2O3 offsets from the new 4D relation defined by stellar mass, N2O3, molecular-gas mass and surface density of stellar mass in comparison with the broad distribution of offsets from the molecular-gas FMR.
N2O3 as measured by the emission-line diagnostic is clearly much better correlated with the predictions from the 4D fundamental relation than it is with those from the molecular-gas FMR.
The corresponding N2O3 dispersion of the 4D relation is 0.034 dex, which is $\sim$ 50\% of the dispersion of the molecular-gas FMR (0.07 dex) in N2O3 axis.

\subsection{Error budget}
In principle, offset measurements in N2O3 axis around the 4D relation can be affected by the observational uncertainties on N2O3, stellar mass, molecular-gas mass and surface density of stellar mass regardless of the projection of the orthogonal distance (this paper) or offset measured along only N2O3 axis (adopted in previous studies).
We here investigate how these four parameters affect on the N2O3 offset measurement.

The observational error of N2O3, $\delta$N2O3, is simply expressed as $\delta$N2O3 =  $[(\delta\log$ [N~{\sc ii}]$\lambda6584)^{2}+(\delta\log$ H$\alpha)^{2}+(\delta\log$ [O~{\sc iii}]$\lambda 5007)^{2}+(\delta\log$ H$\beta)^{2}]^{1/2}$.
Figure \ref{fig6} is the normalised histogram of $\delta$N2O3 in our sample.
The histogram peaks at $\delta$N2O3$\sim$0.03 dex and declines around $\delta$N2O3 $\sim$ 0.06 which roughly corresponds to the selection limit of the sample.
As mentioned in section \ref{sample_selection}, we imposed S/N $>$ 8 for all of [O~{\sc ii}]$\lambda$3726, 3729, H$\beta$, [O~{\sc iii}]$\lambda$5007, H$\alpha$, [N~{\sc ii}]$\lambda$6584 and [S~{\sc ii}]$\lambda$6717, 6731 lines on our sample.
Therefore, the selection limit is determined by emission lines with relatively smaller S/N such as [O~{\sc ii}]$\lambda$3726, 3729 rather than N2O3.

The effects of uncertainties of molecular-gas mass and surface density of stellar mass on N2O3 offset are well below the typical value of N2O3 error (0.03 dex) as following arguments.
The observational error of molecular-gas mass is likely dominated by the SFR error because we assume the Schmidt-Kennicutt law to estimate the molecular-gas mass (see APPENDIX A).
The typical S/N of H$\alpha$ flux in our sample is 81.5 which corresponds to $\delta \log$ SFR = 0.0053, where $\delta \log$ SFR is the observational error of $\log$ SFR.
Given the SFR-$\Delta$metallicity relation of star-forming galaxies with the fixed stellar mass of $\log$ $M_{*}$ = 10 \citep{Mannucci2010}, the uncertainty of 0.0053 dex in $\log$ SFR roughly corresponds to 0.003 dex uncertainty in N2O3.
Based on the similar argument, the possible effect of the error of surface density of stellar mass is roughly estimated to be 0.003-0.005 dex in N2O3 according to the surface-density-$\Delta$N2O3 relation at the fixed stellar mass and molecular-gas mass as shown in Figure \ref{fig3}a.
In this calculation, we assumed that the uncertainty of the surface density of stellar mass is dominated by stellar-mass error which is typically $\delta$ $\log$ $M_{*}$ = 0.06 - 0.092 in our sample depending on stellar mass according to the error estimates provided by MPA/JHU catalogue).
These values (0.003 and 0.003-0.005 dex) are well below the typical observational error of N2O3 (0.03 dex).
Therefore the effects by uncertainties of SFR and surface density of stellar mass are negligible.
The expected N2O3 dispersions (from errors) of both 3D and 4D surfaces are dominated by mass and N2O3 errors.

We tested how the stellar-mass uncertainty affects on the N2O3 offset measurements in the standardised stellar-mass-N2O3 space.
Figure \ref{fig7} is the schematic diagram to show how the offset distances behave when the stellar-mass error changes.
In Figure \ref{fig7}, suppose that the true data point (red dot) on the stellar-mass-N2O3 relation can be observed below the relation (black dot) according to the mass error.
In the top figure, two N2O3 offsets (red and blue dashed lines) measured by two different ways are within the observational uncertainty of N2O3.
However, actual situation in our sample is more similar to the bottom figure in which the stellar-mass error corresponds to 0.09 dex in physical scale.
The red and blue distances can be comparable to or beyond the error of N2O3 depending on how large the mass uncertainty is.
In this sense we can not directly compare the observational N2O3 error (0.03 dex) with measured dispersion in N2O3 regardless of these two different ways to measure the offset.

We therefore performed a simple Monte-Carlo simulation to estimate the expected N2O3 dispersion when we assume typical observational errors of N2O3 (0.03 dex) and stellar mass (0.06 - 0.92 dex depending on stellar mass).
We generated 1000 true data points (like a red dot in Figure \ref{fig7}) following the best-fit function of the stellar-mass-N2O3 relation (black solid line in Figure \ref{fig7}).
At each point on the relation, 1000 artificial observed points are generated which follow the probability distribution function of 2D gaussian with uncertainties of N2O3 and stellar mass (like a black dot in Figure \ref{fig7}.
Thus, we simulated in total $10^{6}$ N2O3 projections of the orthogonal distance distributed around the stellar-mass-N2O3 relation.

\begin{figure*}
	\includegraphics[width=15cm]{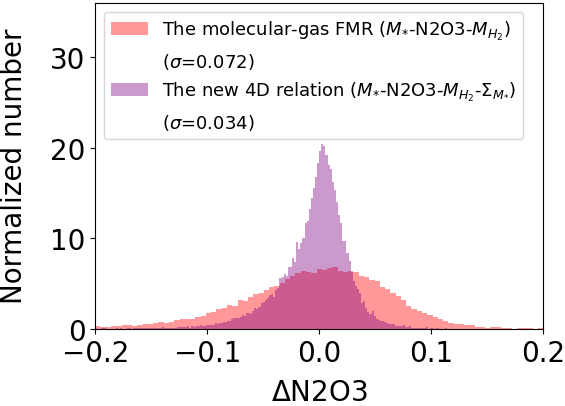}
    \caption{
    Normalised histograms of N2O3 offsets from the molecular-gas FMR and new 4D relation.
    Orthogonal distances from these two relations are projected on the N2O3 axis in the physical scale to calculate the N2O3 offset.
    }
    \label{fig5}
\end{figure*}

\begin{figure*}
	\includegraphics[width=15cm]{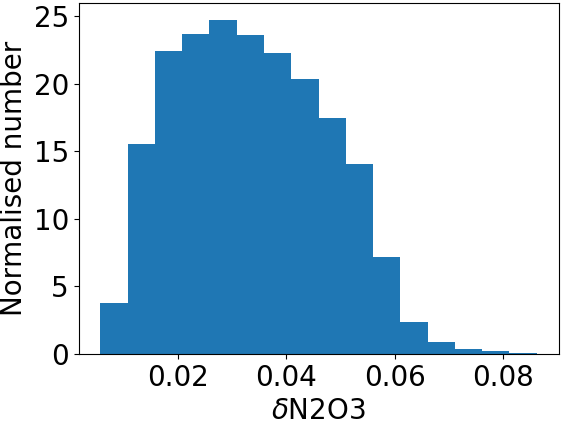}
    \caption{
    Normalised histogram of observational errors of N2O3 in our sample.
    }
    \label{fig6}
\end{figure*}

\begin{figure*}
	\includegraphics[width=15cm]{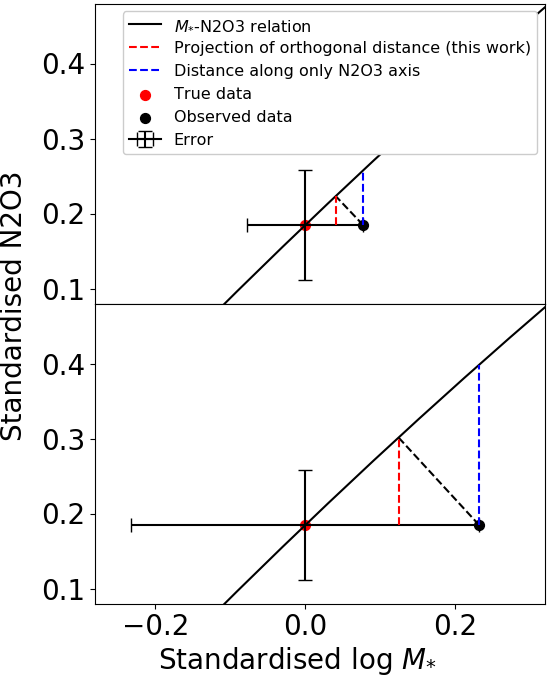}
    \caption{
    Schematic diagrams of offset measurements in N2O3 axis in cases of small (top) and large (bottom) uncertainties on stellar mass.
    Red and blue dashed lines show projection of orthogonal distance (this work) and offset measured along only N2O3 axis (previous works).
    True data point (red dot) on the stellar-mass-N2O3 relation (black solid line) can be observed below the relation according to the observational errors of N2O3 and stellar mass.
    Note that both axes are standardised.
    N2O3 error corresponds to 0.03 dex in physical scale.
    Stellar-mass errors in top and bottom correspond to 0.03 and 0.09 dex in physical scale, respectively.
    }
    \label{fig7}
\end{figure*}

\begin{figure*}
	\includegraphics[width=15cm]{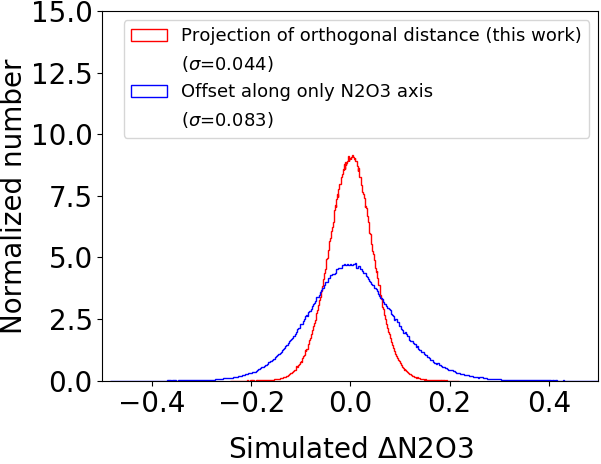}
    \caption{
    N2O3 dispersions in physical scale expected from typical observational errors of N2O3 and stellar mass based on a Monte-Carlo simulation.
    Red and blue are histograms of N2O3 projections of orthogonal distances and offsets along only N2O3 axis, respectively.
    }
    \label{fig8}
\end{figure*}

The standard deviation of the simulated N2O3 projections of orthogonal distances is $\sim$0.04 dex in N2O3 physical scale as shown in Figure \ref{fig8}, i.e. we expect $\sim 0.04$ dex dispersion in N2O3 due to the error budgets of N2O3 and stellar mass.
The actual dispersion of N2O3 projections of orthogonal distances around the 3D relation (0.07 dex in N2O3 shown in Figure \ref{fig5}) is larger than $\sim$0.04 dex, therefore, suggesting a need for the new parameter.
This dispersion can be further reduced by introducing a new parameter.
Indeed it was reduced to 0.034 dex in our 4D analysis in Figure \ref{fig5}.
The expected N2O3 dispersion calculated by Monte-Carlo simulation ($\sim$0.04 dex) is comparable to the actual dispersion of N2O3 projections of orthogonal distances around the new 4D relation (0.034 dex), which indicates the dispersion around the 4D relation is dominated by the observational errors of N2O3 and stellar mass.

In summary, the new fundamental metallicity relation including the surface density of stellar mass defines a tighter hypersurface in the 4D parameter space.

\section{Discussion}
\label{discussion}
\subsection{Radial metallicity gradient}
\begin{figure*}
	\includegraphics[width=15cm]{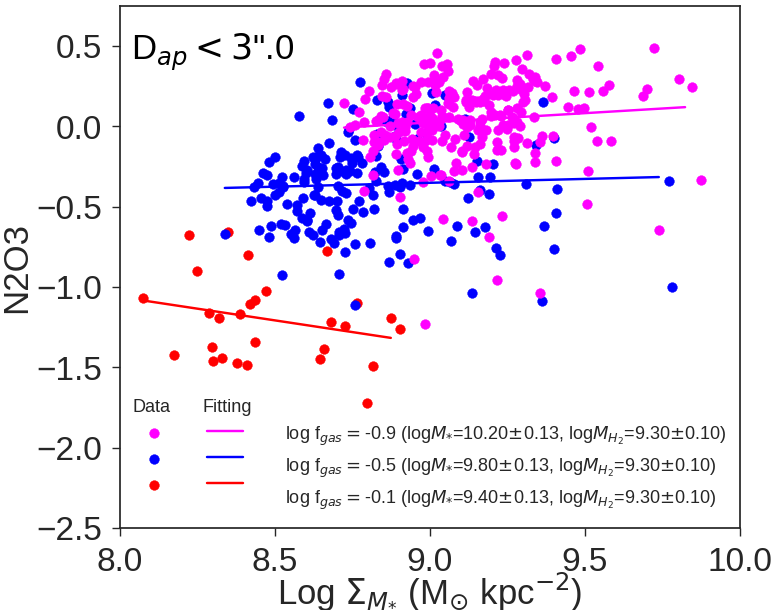}
    \caption{
    Relations between surface density of stellar mass and metallicity based on apparently small galaxies.
    Galaxies appearing smaller than the aperture size of SDSS spectroscopic data (3$\arcsec$.0 in diameter) are selected to remove the possible aperture effect on metallicity.
    The binning size is slightly larger than that in Figure \ref{fig3} to ensure a sufficient sample size.
    Solid lines indicate the best fits of linear functions for each binned sample.
    }
    \label{fig9}
\end{figure*}

A radial gradient of metallicity within a galaxy could cause metallicity to depend on surface density of stellar mass.
For example, metallicity is generally lower in the outskirts of a spiral galaxy \citep{Zaritsky1994}.
We measured the metallicity of our sample within a fixed SDSS aperture size (3$\arcsec$.0 in diameter) to trace more of the outer part for smaller galaxies while concentrating on the central part for larger galaxies.
A smaller size of a galaxy corresponds to a higher surface density for a fixed stellar mass. 
The negative correlation of high-$f_{gas}$ galaxies shown in Figure \ref{fig3} and \ref{fig4} could be an aperture effect, but such an effect would probably have little impact for two reasons.
Firstly, the measurement of metallicity is, in general, flux weighted and biased towards the central bright region regardless of the size of the galaxy.
Secondly, a lower redshift limit ($z > 0.04$) is imposed on the sample to ensure that the metallicity is approximately the same as the global value \citep{Kewley2005}.
We tested the surface-density-N2O3 relation based on galaxies that appeared smaller than the spectroscopic aperture size to limit substantially any possible aperture effect. 
Although this apparent-size limit traces only high-surface-density galaxies and so the resulting sample is small, the general trend of the surface-density-N2O3 relations remains the same (Figure \ref{fig9}).
Hence, the surface-density-N2O3 relations are unlikely to be attributable to any aperture effect.

\subsection{Outflow and inflow}
Galactic outflow and inflow are possible candidates for the link between metallicity and surface density of stellar mass.
Enriched gas outflow can remove metals from a galaxy, and this process is more effective in the shallower gravitational potentials of less massive galaxies if the wind velocity is roughly constant over the whole range of stellar mass, resulting in the well-known stellar-mass-metallicity relation \citep{Garnett2002,Tremonti2004}.
Suppose that two galaxies with the same stellar mass have almost the same depth of gravitational well, but that the surface gravity depends on the galactic surface density.
Galactic wind might easily escape in the case of less concentrated surface gravity on average, thereby giving rise to the surface-density-N2O3 relation.
However, cosmological hydrodynamic simulations suggest that such a constant-velocity model disagrees strongly with the observed slope and scatter of the stellar-mass-metallicity relation and that the wind velocity is always high enough to escape the gravitational well \citep{Finlator2008}.
In such a case, the N2O3 dependence of surface gravity should disappear.

Metallicity dilution and star formation caused by inflow of metal-poor gas is another possible candidate.
If the dilution time scale is longer than the dynamical time scale of infall (i.e. $t_{dil} > t_{dyn}$), the galaxy is in the process of returning to the equilibrium metallicity as determined by the dilution and chemical enrichment by newly invoked star formation.
Such a galaxy might have observationally lower metallicity compared with the equilibrium metallicity because the bulk of emission lines is emitted from the star-forming regions in metal-poor infall gas. 
Because the ratio $t_{dil}/t_{dyn}$ is proportional to $\sqrt \rho$, where $\rho$ is the mass density \citep{Ellison2008}, galaxies with higher surface mass density may be expected to have larger values of $t_{dil}/t_{dyn}$, suggesting lower metallicity.
This time lag before metallicity dilution is expected to be more evident in gas-rich galaxies that are undergoing active infall and star formation.
This is reflected exactly in the negative correlation for high-$f_{gas}$ galaxies shown in Figure \ref{fig4}, whereas the low-$f_{gas}$ galaxies indicate the opposite trend of this metallicity-dilution scenario.
The latter type of galaxy is probably in the terminal stage of molecular-gas depletion, which means that its metallicity is already in equilibrium.

\subsection{Efficiency}
Star-formation efficiency is a likely reasonable explanation for the surface-density-N2O3 relation of low-$f_{gas}$ galaxies.
The Schmidt-Kennicutt law indicates that a more concentrated galaxy has a higher SFR compared with the molecular-gas mass, in other words a higher SFR/$M_{H_{2}}$ ratio (see APPENDIX A).
This means that galaxies with higher surface density are more efficient at turning gas into stars, resulting in higher metallicity.
This expected trend is consistent with the observed surface-density-N2O3 relation of low-$f_{gas}$ galaxies shown in Figure \ref{fig4}.
Star-formation efficiency has been discussed as an alternative to the aforementioned galactic-wind scenario as a means to reproduce the stellar-mass-metallicity relation \citep{Brooks2007,Dalcanton2007,Calura2009}.
Massive galaxies form stars more effectively and are more chemically evolved than do less massive galaxies, which is known as the \lq downsizing\rq formation of galaxies. 
Such downsizing is related to the total stellar mass of the galaxy.
The surface-density-N2O3 relation suggests that galaxies with higher surface density form stars more effectively and are more chemically evolved than do galaxies with lower surface density for a fixed stellar mass. 

\subsection{Global and local physical parameters}
The dilution time scale of gas inflow and the star-formation efficiency could be, to a greater or lesser extent, underlying mechanisms that are common to all star-forming galaxies.
The dependence of N2O3 on surface density of stellar mass is probably regulated by a combination of these two mechanisms.
The fractional contribution of gas inflow could be greater in high-$f_{gas}$ galaxies, whereas the star-formation efficiency is likely more important in low-$f_{gas}$ galaxies.
The relatively gentle slope of galaxies in the mid-$f_{gas}$ bin in Figure \ref{fig4} may be due to the relatively weak contribution of the gas inflow.

The physical mechanism of SFR and surface density to affect metallicity could be different, although the impact of SFR on the stellar-mass metallicity relation is still in debate \citep{Telford2016,Kashino2016,Sanchez2017,Barrera-Ballesteros2017}.
Thus both may exist.
However, it is not a focus of our paper to rule out the re-examined results that claim a lack or weaker dependence of SFR on MZR.

The second-order effects of surface density on metallicity likely originate from the local physical conditions of a galaxy.
It has been reported recently that the locally defined fundamental relations of star-forming galaxies can, to some extent, reproduce the global properties of the stellar-mass-metallicity relation and the radial metallicity gradient simultaneously by integrating spatially resolved data \citep{Barrera-Ballesteros2016}.
The main implication of this is that the local physical properties are closely involved in determining the global metallicity property.
The surface density of a galaxy is likely a bridge between such local and global natures.
Our findings indicate that it is not sufficient to consider only global relations such as the FMR and molecular-gas FMR.
Both global and local parameters are needed to understand the physics of star-forming galaxies.
The representative metallicity of a galaxy is regulated by not only global parameters but also local physics such as dilution and star-formation efficiency as traced by the surface density.
The surface density of a galaxy is one of the most basic observables, and it will be easily measured in upcoming large imaging surveys such as the Dark Energy Survey \citep{DES2005}, Hyper Suprime-Cam \citep{HSC2012} and the Large Synoptic Survey Telescope \citep{LSST2008}.
We suggest that future FMR studies consider the FMR in 4D space, including the surface density of stellar mass.

%\begin{figure}
%	\includegraphics[width=\columnwidth]{example}
%    \caption{This is an example figure. Captions appear below each figure.
%	Give enough detail for the reader to understand what they're looking at,
%	but leave detailed discussion to the main body of the text.}
%    \label{fig:example_figure}
%\end{figure}

%\begin{table}
%	\centering
%	\caption{This is an example table. Captions appear above each table.
%	Remember to define the quantities, symbols and units used.}
%	\label{tab:example_table}
%	\begin{tabular}{lccr} % four columns, alignment for each
%		\hline
%		A & B & C & D\\
%		\hline
%		1 & 2 & 3 & 4\\
%		2 & 4 & 6 & 8\\
%		3 & 5 & 7 & 9\\
%		\hline
%	\end{tabular}
%\end{table}

\section{Conclusions}
\label{conclusion}
We performed PCA on 41,338 star-forming galaxies selected from SDSS Data Release 7. 
In six groups of physical parameters, stellar mass, N2O3, molecular-gas mass, surface density of stellar mass, dark-matter halo mass and ionization parameter are found to show the highest values of factor loadings in the 3D space of PC1-3.
Among the combinations of these parameters, the dispersion around the fundamental relation defined by stellar mass, N2O3, molecular-gas mass and surface density of stellar mass is the smallest, which indicates that the fourth most important parameter of star-forming galaxies is the surface density of stellar mass.
The reduction of the dispersion by adding the fourth parameter is predominant in N2O3 axis, i.e., the fourth parameter reduces N2O3 dispersion to 50\% of that of the molecular-gas FMR.
We found that the surface-density-N2O3 relation depending on gas-mass fraction is responsible for the reduction of the dispersion around the fundamental relation.
The dilution time scale of gas inflow in gas-rich galaxies and the star-formation efficiency in gas-poor galaxies could explain the observational dependence on surface density.
We suggest that future FMR studies consider the FMR in 4D space, including the surface density of stellar mass.

\section*{Acknowledgements}
%The authors thank the anonymous referees for many insightful comments.
We thank the anonymous referees for many insightful comments.
We are very grateful to Teimoorinia Hossen and Ellison Sara L. for providing a catalogue of physical parameters of SDSS star-forming galaxies. 
T.G. acknowledges the support by the Ministry of Science and Technology of Taiwan through grant NSC 103-2112-M-007-002-MY3, and 105-2112-M-007-003-MY3.

%%%%%%%%%%%%%%%%%%%%%%%%%%%%%%%%%%%%%%%%%%%%%%%%%%



\bibliographystyle{mnras}
\bibliography{mnras_FMR} % if your bibtex file is called example.bib

%\begin{thebibliography}{99}
%\end{thebibliography}

%%%%%%%%%%%%%%%%%%%%%%%%%%%%%%%%%%%%%%%%%%%%%%%%%%

%%%%%%%%%%%%%%%%% APPENDICES %%%%%%%%%%%%%%%%%%%%%
\appendix

\section{Physical parameters of star-forming galaxies}
In total, 18 physical parameters of 41,338 galaxies were compiled from various literature sources, and a further 11 were calculated in this work.
Logarithms in physical parameters were taken before standardization.
We summarise the details of all 29 parameters below.\\

\begin{itemize}
\item $M_{*}$ (M$_{\odot}$), Stellar mass was collected from the publicly available MPA/JHU catalogue, in which galaxy-population-synthesis models \citep{Bruzual2003} are fitted to multi-band photometric data to estimate stellar mass.
Standardised log $M_{*}$ was used in the PCA.
\item $M_{virial}$ (M$_{\odot}$), Virial mass was estimated as 2$r_{half}\sigma^{2}/G$. 
$r_{half}$ and $\sigma$ are $r$-band half-light radius and rest-frame velocity dispersion, respectively.
G is gravitational constant. 
Standardised log $M_{virial}$ was used in the PCA.
\item $\sigma$ (km s$^{-1}$), Rest-frame velocity dispersion was collected from the MPA/JHU catalogue.
Standardised log $\sigma$ was used in the PCA.
\item $M_{igas}$ (M$_{\odot}$), Ionised-gas mass was estimated as \\
$2.33 \times 10^{3}(L_{H_{\alpha}}/10^{39}erg s^{-1})(10^{3}cm^{-3}/n_{e})$ M$_{\odot}$ \\
under the assumption of case B recombination \citep{Osterbrock1989}. 
$L_{H_{\alpha}}$ and $n_{e}$ are H$\alpha$ luminosity and electron density, respectively.
Standardised log $M_{igas}$ was used in the PCA.
\item M$_{u}$, M$_{g}$, M$_{r}$, M$_{i}$, M$_{z}$, Absolute magnitudes in $u$, $g$, $r$, $i$ and $z$ bands as collected from the literature \citep{Teimoorinia2017}.
Standardised M$_{u}$, M$_{g}$, M$_{r}$, M$_{i}$ and M$_{z}$ were used in the PCA.
\item N2O3, Metallicity indicator was calculated as N2O3 $=\log$[([N~{\sc ii}]$\lambda 6584/{\rm H}\alpha$)/([O~{\sc iii}]$\lambda 5007/{\rm H}\beta$)].
N2O3 can be converted to metallicity through $12+\log({\rm O/H})=8.73+0.32 \times {\rm N2O3}$ based on the PP04O3N2 method with an uncertainty of $\sim$ 0.14 dex \citep{Pettini2004}.
In addition, 14 other metallicity calibrations \citep{Denicolo2002,Zaritsky1994,McGaugh1991,Pettini2004,Pilyugin2010,Maiolino2008,Marino2013,Kewley2002,Kobulnicky2004,Kewley2008} were examined to confirm the universality of the surface-density-metallicity relation by using a metallicity calculation code \citep{Bianco2016}. 
The same trend of the surface-density-metallicity relation shown in Figure \ref{fig4} comes out in all of these metallicity calibrations.
Standardised N2O3 was used in the PCA.
\item $M_{metal}$ (M$_{\odot}$), Metal mass in ionised gas was calculated as $10^{12+\log({\rm O/H})-8.69} \times 0.0181 \times {M}_{igas}$, where the factors of 8.69 and 0.0181 are the solar oxygen abundance and metal mass ratio, respectively \citep{Asplund2009}. 
Standardised log $M_{metal}$ was used in the PCA.
\item A$_{V}$, V-band dust extinction was calculated from the observed Balmer decrement (H$\alpha$/H$\beta$) assuming the extinction law of star-forming galaxies \citep{Calzetti2000}.
Standardised A$_{V}$ was used in the PCA.
\item SFR (M$_{\odot}$ yr$^{-1}$), Star-formation rate was collected from the MPA/JHU catalogue, in which SFR is calculated from H$\alpha$ luminosity corrected for dust extinction and aperture loss  \citep{Brinchmann2004}.
Standardised log SFR was used in the PCA.
\item sSFR (yr$^{-1}$), Specific SFR was calculated as SFR/M$_{*}$.
Standardised log sSFR was used in the PCA.
\item EW$_{{\rm H}\alpha}$ (\AA), Rest-frame equivalent width was collected from the MPA/JHU catalogue. 
Standardised log EW$_{{\rm H}\alpha}$ was used in the PCA.
\item D4000, Index of 4000-\AA\  break was collected from the MPA/JHU catalogue.
D4000 is defined as $\langle F^{+}\rangle/\langle F^{-}\rangle$, where $\langle F^{+}\rangle$ and $\langle F^{-}\rangle$ are flux densities averaged over 4050-4250 \AA\ and 3750-3950 \AA\ \citep{Bruzual1983}.
Standardised log D4000 was used in the PCA.
\item $g-r$, Colour of $g-r$ was collected from the literature \citep{Teimoorinia2017}.
Standardised $g-r$ was used in the PCA.
\item $M_{\rm HI}$, Atomic hydrogen gas mass was collected from the literature \citep{Teimoorinia2017}. 
It has been reported that $M_{\rm HI}$ can be estimated empirically from the combination of 15 physical parameters of a star-forming galaxy \citep{Teimoorinia2017}, although the physical mechanism underlying this relation is still open to debate.
Standardised log $M_{\rm HI}$ was used in the PCA.
\item $M_{\rm H_{2}}$, Molecular-gas mass was estimated assuming the Schmidt-Kennicutt law \citep{Kennicutt1998}, which is an empirical relation between surface density of SFR and molecular-gas mass, i.e. $\Sigma_{\rm SFR}=1.6\times10^{-4}(\Sigma_{\rm gas}/M_{\odot} pc^{-2})^{1.4} M_{\odot}$ yr$^{-1}$ kpc$^{-2}$.
This relation is equivalent to $M_{gas} \propto ({\rm SFR})^{5/7}(\pi r^{2})^{2/7}$.
Therefore, the molecular-gas mass can be calculated from the SFR and the size of the galaxy by assuming this relation.
The Schmidt-Kennicutt law can also be expressed as SFR/$M_{gas}$ $\propto$ $(\Sigma_{\rm gas})^{0.4} = (f_{gas} \Sigma_{\rm M_{*}})^{0.4}$, which indicates a higher star-formation efficiency in a more concentrated galaxy. 
Standardised log $M_{\rm H_{2}}$ was used in the PCA.
Although the derivations of the above two gas masses are not direct methods, they are included in the analyses to estimate the importance of gas mass.
\item $r_{disk}$ (kpc), Exponential disk scale length was collected from the literature \citep{Teimoorinia2017}.
Galaxy structural parameters were measured from bulge+disk decompositions, i.e. a pure exponential disk and de Vaucouleurs bulge \citep{Simard2011}.
Standardised log $r_{disk}$ was used in the PCA.
\item $r_{half}$ (kpc), Half-light radius along a semi-major axis in r-band \citep{Simard2011} was collected from the literature \citep{Teimoorinia2017}.
Standardised log $r_{half}$ was used in the PCA.
\item $\Sigma_{M_{*}}$ (M$_{\odot}$ kpc$^{-2}$), Surface density of stellar mass was calculated as M$_{*}$/$\pi r_{half}^{2}$.
Standardised log $\Sigma_{M_{*}}$ was used in the PCA.
\item $\Sigma_{\rm SFR}$ (M$_{\odot}$ yr $^{-1}$kpc$^{-2}$), Surface density of SFR was calculated as SFR/$\pi r_{half}^{2}$.
Standardised log $\Sigma_{\rm SFR}$ was used in the PCA.
\item B/T, Bulge-to-total fraction in $r$-band was collected from the literature \citep{Teimoorinia2017}. 
This fraction is based on a pure exponential disk and de Vaucouleurs bulge decompositions \citep{Simard2011}.
Standardised B/T was used in the PCA.
\item $\delta_{5}$, Local galaxy number density was collected from the literature \citep{Teimoorinia2017} and is defined as $\sum_{n}=n/\pi d_{n}^{2}$ (here, $n$=5).
$d_{n}$ is the projected distance in Mpc to the $n$-th nearest neighbour within $\pm$ 1000km s$^{-1}$. 
$\delta_{5}$ is normalized by the median $\sum_{n}$ within a redshift slice of $\pm$ 0.01.
Standardised log $\delta_{5}$ was used in the PCA.
\item $M_{halo}$ (M$_{\odot}$), Mass of dark-matter halo hosting galaxies was collected from the literature \citep{Teimoorinia2017}.
The halos were identified in previous work \citep{Yang2007,Yang2009}.
Standardised log $M_{halo}$ was used in the PCA.
\item $n_{e}$ (cm$^{-3}$), Electron density was calculated from the line ratio of [S~{\sc ii}]$\lambda6717$/[S~{\sc ii}]$\lambda6731$ assuming an electron temperature of $10^{4}$ K. 
In cases in which the ratio was beyond the theoretical asymptotic values \citep{Osterbrock1989}, [O~{\sc ii}]$\lambda$3729/[O~{\sc ii}]$\lambda$3726 was used instead. 
If both were out of range, [S~{\sc ii}]$\lambda 6717$/[S~{\sc ii}]$\lambda 6731$ was replaced with the asymptotic values of $\sim$1.4 or 0.5.
Standardised log $n_{e}$ was used in the PCA.
\item $q$, Indicator of ionisation parameter was calculated as [O~{\sc iii}]$\lambda$5007/[O~{\sc ii}]$\lambda$3727 corrected for dust extinction.
Standardised log [O~{\sc iii}]$\lambda$5007/[O~{\sc ii}]$\lambda$3727 was used in the PCA.
\item $z$, Redshift was collected from the MPA/JHU catalogue.
Standardised log $z$ was used in the PCA.
\end{itemize}

\section{Factor loading of PCA}
Because various measurements with different units are used simultaneously in the PCA, all data were standardised for a fair comparison.
The standardised parameter X$^{i}_{std}$ is defined as X$^{i}_{std}$=(X$^{i}$-X$_{ave}$)/$\sigma_{\rm X}$, where X$^{i}$ is an original parameter of the $i$-th galaxy.
$X_{ave}$ and $\sigma_{X}$ are the average and standard deviation, respectively, of parameter X.

PCA is a linear transformation of the original parameters.
The vector of the new axes, namely, $\vec{\rm PC}=({\rm PC1}, {\rm PC2}, ..., {\rm PC}n)$, can be expressed as $\vec{\rm PC}=A\cdot \vec{\rm X}$, where $\vec{\rm X}$ is the vector of original parameters.
The fractional contributions of original parameters to each new axis PC$n$ are $\vec{a_{n}}$=(a$_{n1}$, a$_{n2}$, ..., a$_{nn}$), which are the $n$-th-row components of the matrix $A$.
The contribution of the PC$n$ axis to the total dispersion is proportional to the eigenvalue of PC$n$ ($l_{n}$).
Therefore, the factor loadings defined as $\sqrt[]{\mathstrut l_{n}}\vec{a_{n}}$ indicate the contribution of each original parameter to the total dispersion.

%%%%%%%%%%%%%%%%%%%%%%%%%%%%%%%%%%%%%%%%%%%%%%%%%%

% Don't change these lines
\bsp	% typesetting comment
\label{lastpage}
\end{document}